\documentclass{aa}  
\usepackage{graphicx}
\usepackage{float}
\usepackage{xcolor}
\usepackage{txfonts}
\usepackage[colorlinks,allcolors={blue}]{hyperref}
\usepackage{lineno}
\usepackage{ulem}
\definecolor{grey}{rgb}{0.,0.5,0.}
\definecolor{darkblue}{rgb}{0,0,0.5}
\definecolor{orange}{rgb}{1.,0.5,0.}
\definecolor{darkgreen}{rgb}{0,0.5,0}

\newcommand{\sect}[1]{\text{Section~\ref{#1}}}
\newcommand{\fig}[1]{\text{Figure~\ref{#1}}}

\newcommand{\app}[1]{\text{Appendix~\ref{#1}}}

\newcommand{\stagger}{\texttt{STAGGER}}

\newcommand{\marcs}{\texttt{MARCS}}
\newcommand{\balder}{\texttt{Balder}}

\newcommand{\kms}{\mathrm{km\,s^{-1}}}
\newcommand{\ms}{\mathrm{m\,s^{-1}}}
\newcommand{\nm}{\mathrm{nm}}
\newcommand{\teff}{T_{\mathrm{eff}}}
\newcommand{\logg}{\log{g}}
\newcommand{\feh}{\mathrm{[Fe/H]}}
\newcommand{\cafe}{\mathrm{[Ca/Fe]}}
\newcommand{\dex}{\mathrm{dex}}
\newcommand{\ccf}{\mathcal{CCF}}

\newcommand{\vmic}{\xi_{\mathrm{mic.}}}
\newcommand{\hk}{H$\&$K}
\newcommand{\lgeps}[1]{A(\mathrm{#1})}
\definecolor{cis}{rgb}{0.78, 0.08, 0.52}

\begin{document} 

   %\title{A spectral grid of the \ion{Ca}{II} triplet in 3D non-LTE: abundance corrections and radial velocities for metal-poor stars.}
   \title{3D non-LTE \ion{Ca}{II} line formation in metal-poor FGK stars}
   \subtitle{I. Abundance corrections, radial velocity corrections, and synthetic spectra}

  %\authorrunning{C. Lagae et al.}
  \titlerunning{3D non-LTE formation of \ion{Ca}{II} lines.}
  
   \author{C.\,Lagae\inst{\ref{SU}},
          A.\,M.\,Amarsi\inst{\ref{UU}},
          K.\,Lind\inst{\ref{SU}},
            }

   \institute{\label{SU}Department of Astronomy, Stockholm University,
              Albanova University Center, Roslagstullsbacken 21, 106 91 Stockholm, Sweden\\
              \email{cis.lagae@astro.su.se}
              \and 
              \label{UU}Theoretical Astrophysics,
              Department of Physics and Astronomy, Uppsala University, 
              Box 516, SE-751 20 Uppsala, Sweden
             }

   \date{Accepted 13 March 2025; Received 4 Nov 2024}

% \abstract{}{}{}{}{} 
% 5 {} token are mandatory
 
  \abstract
  % context heading (optional)
  % {} leave it empty if necessary  
   {The \ion{Ca}{II} near-ultraviolet resonance doublet (\hk{}) and the near-infrared triplet (CaT) are among the strongest features in stellar spectra of FGK-type stars. These spectral lines remain prominent down to extremely low metallicities and are thus useful for providing stellar parameters via ionisation balance, for Galactic chemical evolution, and as radial velocity diagnostics. However, the majority of studies that model these lines in late-type stars still rely on simplified one dimensional (1D) hydrostatic model atmospheres and the assumption of local thermodynamic equilibrium (LTE).}
  % aims heading (mandatory)
   {We present 3D non-LTE radiative transfer calculations of the CaT and \hk{} lines in an extended grid of 3D model atmospheres of metal-poor FGK-type. We investigate the impact of 3D non-LTE effects on abundances, line bisectors and radial velocities.}
  % methods heading (mandatory)
   {We used a subset of 3D model atmospheres from the recently published \stagger{}-grid to synthesize spectra in 3D (non-)LTE with \balder{} for nine different calcium-to-iron ratios. For comparison, similar calculations were performed in 1D (non-)LTE using models from the \marcs{} grid.}
% results heading (mandatory)
   {Abundance corrections for the CaT lines relative to 1D LTE range from $+0.1\gtrsim\Delta^\mathrm{3N}_\mathrm{1L}\gtrsim-1.0~\dex$, with more severe corrections for strong lines in giants. With fixed line strength, the abundance corrections become more negative with increasing effective temperature and decreasing surface gravity. Radial velocity corrections relative to 1D LTE based on cross-correlation of the whole line profile range from $-0.2~\kms$ to $+1.5~\kms$, with more severe corrections where the CaT lines are strongest. The corrections are even more severe if the line core alone is used to infer the radial velocity.}
% conclusions heading (optional), leave it empty if necessary 
   {The line strengths and shapes, and consequently the abundance and radial velocity corrections, are strongly affected by the chosen radiative transfer assumption, 1/3D (non)-LTE. We release grids of theoretical spectra that can be used to improve the accuracy of stellar spectroscopic analyses based on the \ion{Ca}{II} triplet lines.}


   \maketitle
%
%-------------------------------------------------------------------

\section{Introduction}\label{Sec:introduction}

The oldest, metal-poor, stars provide an opportunity to study the early chemical and dynamical evolution of the Milky Way and its satellites, and that of the different stellar populations therein.  The detailed abundance patterns of metal-poor stars can be combined with kinematical information to distinguish between different structures in the Milky Way and even identify historical merger products such as Gaia-Enceladus \citep{Belokurov18,Haywood18,Helmi18}.
Moreover, extremely metal-poor stars (EMP: $\mathrm{[Fe/H]}<-3$; \citealt{Beers05}) could, in principle, be formed from a cloud enriched by a single first star supernova \citep{Tominaga07b,Nomoto13,Keller14,Frebel15}. Since these first stars, otherwise known as metal-free or Population (Pop) III stars, remain elusive to our observations \citep{Hartwig15}, the chemical composition of EMP stars allow us to indirectly constrain their properties (e.g. \citealt{Placco15} and \citealt{Nordlander17a}). 

Singly-ionised calcium plays a unique and important role in this context: it has several prominent spectral lines that remain observable even at extremely low metallicities. 
Metal-poor stars are a rare occurrence in the solar neighbourhood, with approximately $\sim1$ EMP star in $200\,000$ \citep{Frebel13}. Extensive effort has been made in the last decades to find and characterise metal-poor stars by targeted observational campaigns following the pioneering work of the HK survey \citep{Beers85,Beers92} and Hamburg/ESO Survey \citep{Wisotzki00,Christlieb03,Christlieb08}. In this endeavour, the \ion{Ca}{II} near ultraviolet resonance doublet lines at $397\,\nm$ and $393\,\nm$ (\hk{}) have been used with great success to more efficiently target metal-poor stars in large photometric surveys such as Pristine \citep{Starkenburg17b} and SkyMapper \citep{Keller07}. 

Of similar importance are the \ion{Ca}{II} near-infrared triplet lines at $8497\,\AA$, $8542\,\AA$, and $8662\,\AA$ (CaT). These lines are typically the strongest feature in the near infrared and experience little to no interstellar absorption. Moreover, the stellar flux of K-type red giants peaks in the near-infrared, making these lines an ideal target for large spectroscopic studies (e.g. Gaia \citep{Gaia20,Gaia23}, 4MOST \citep{deJong19}, Weave \citep{Dalton12}). As such, the CaT lines have been used in a variety of applications, ranging from metallicity calibration, abundance analysis, radial velocity determination and constraining the properties of the first stars.

Typically, stellar abundances and radial velocities are determined by comparing synthetic spectra, computed under the assumption of local thermodynamic equilibrium (LTE) using one-dimensional (1D) model atmospheres, to observations. Such 1D LTE synthetic spectra are easy to compute but come at the cost of extra free parameters such as micro- and macro-turbulence, to approximate the physics of stellar sub-surface convection. It has long been known that the assumption of 1D and LTE can severely alter the line shape and strength, leading to under- or overestimation of stellar abundances \citep{Nordlund80,Kiselman1991}.
Moreover, 1D models cannot make any predictions about the observed convective blueshift of spectra lines \citep{Dravins1981}. The fundamental 3D and non-LTE effects are typically stronger at lower metallicity, lower surface gravity, and higher effective temperature \citep[e.g][and references therein]{Lind24}, although different effects can sometimes cancel and give results closer to 1D LTE.

\begin{figure*}
  \centering
  \resizebox{7.in}{!}{\includegraphics{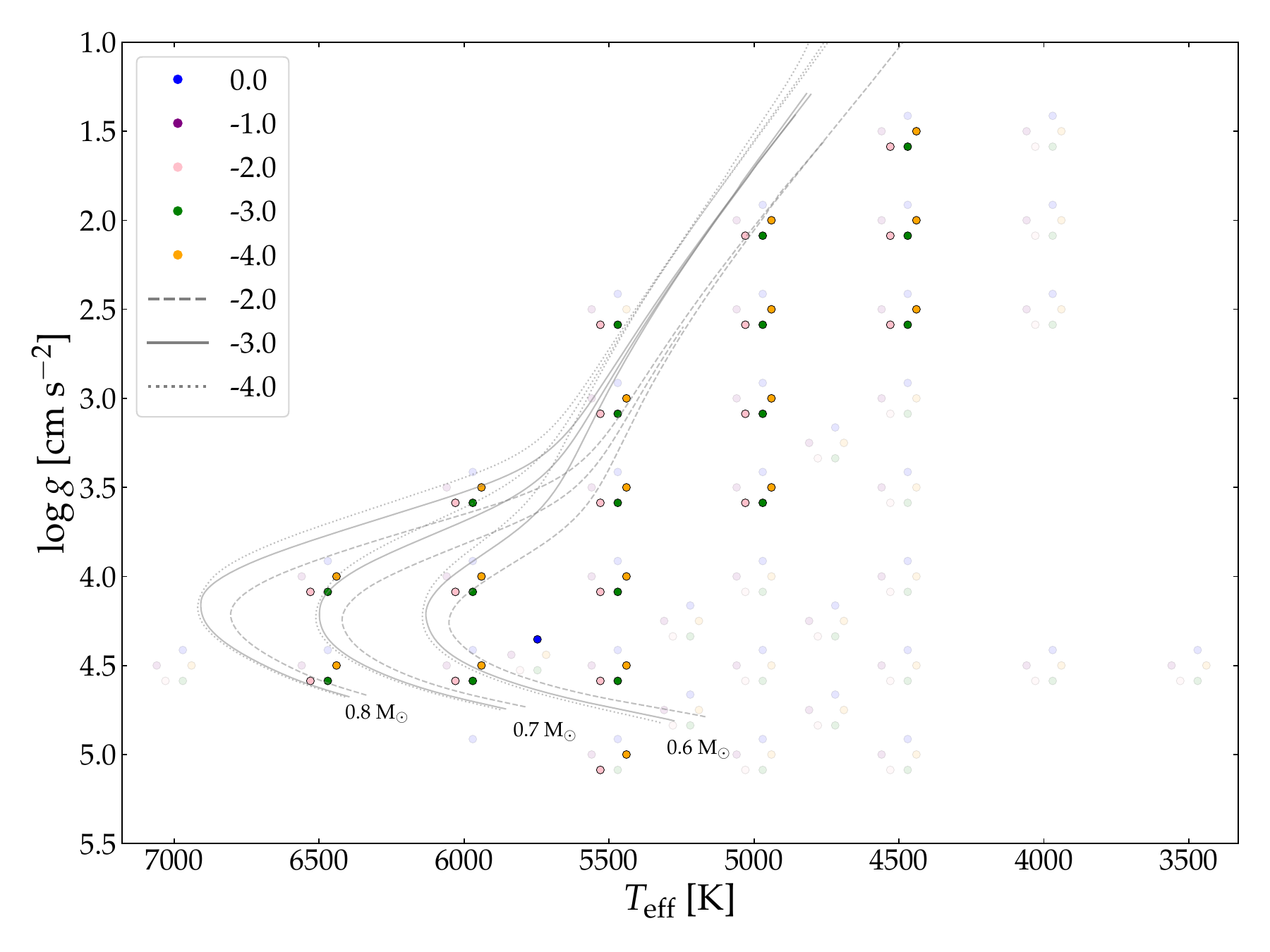}}
\caption[]{Overview of the 3D model grid used in this work in relation to the \stagger{}-grid, limited to solar and metal-poor models. [Fe/H] is colour coded; \stagger{}-grid models not used in this work are transparent. MIST evolutionary tracks are overplotted \citep{Dotter16,Choi16,Paxton11,Paxton13,Paxton15}  for three stars with initial mass of $0.6$, $0.7$ and $0.8~\mathrm{M}_\odot$ at metallicities of $\mathrm{[Fe/H]}=-2,~-3~\mathrm{and}~-4$.}
\label{fig:gridstatus}
\end{figure*}

The first 1D non-LTE study of \ion{Ca}{II} was conducted by \cite{Jorgensen1992}, followed by more recent studies by \cite{Andretta05}, \cite{Mashonkina07,Mashonkina17a}, \cite{Starkenburg10}, \cite{Merle11}, \cite{Spite12}, \cite{Sitnova19} and \cite{Osorio19,Osorio20,Osorio22}. More comprehensive modelling of the CaT lines in 3D LTE or 3D non-LTE has been performed for the Sun \citep{Leenaarts09a,DeLaCruzRodriguez11,Leenaarts14,DeLaCruzRodriguez15,Moe24}, where their line cores form in the chromosphere and can be used as a magnetic activity indicator \citep{Socas-Navarro00a,QuinteroNoda16,Huang24}. On the contrary, there has been only a handful of 3D non-LTE studies that targeted other stars \cite{Lind13,Nordlander17a,Wedemeyer17,Lagae23}. 

\begin{figure}
  \centering
  \resizebox{3.5in}{!}{\includegraphics{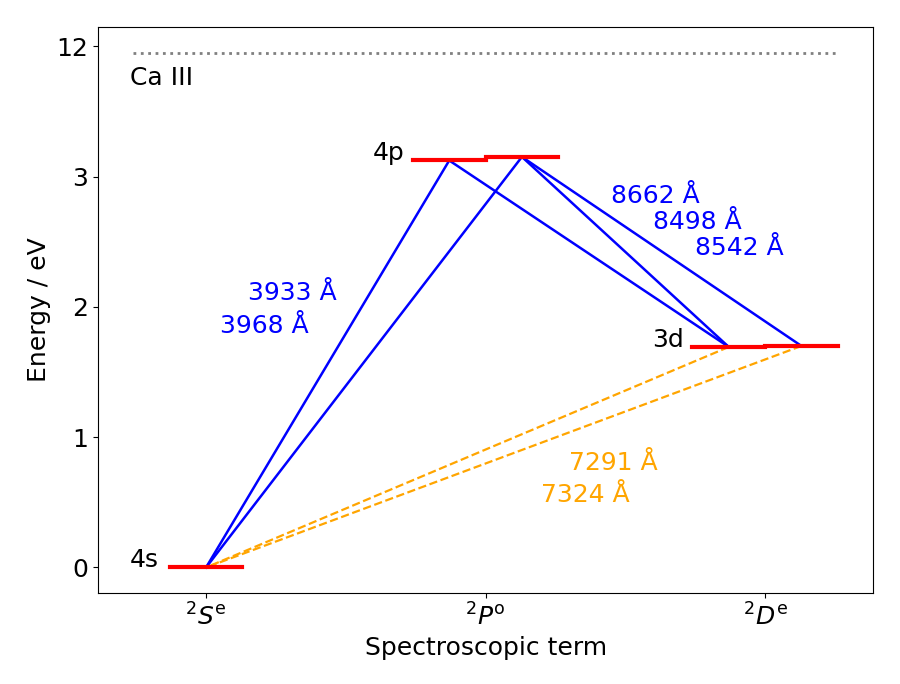}}
\caption[]{Grotrian diagram for the model atom. Allowed bound-bound transitions in solid blue and forbidden radiative bound-bound transitions are shown with dashed orange lines.}
\label{fig:term_diagram}
\end{figure}

Performing 3D non-LTE spectroscopic analyses has become more feasible through developments of grids of 3D model atmospheres \citep[e.g.][]{Freytag12,BertrandeLis22,Rodriguez24}, and larger amounts as well as more accurate atomic data \citep{Barklem16}; 
combined with the accessibility of high-performance computing. To utilise the 3D non-LTE method in upcoming surveys of tens of millions of stars such as 4MOST \citep{deJong19}, PLATO \citep{Rauer14,Rauer24}, and WEAVE \citep{Dalton12}, one approach is to pre-compute grids of synthetic spectra and abundance corrections, which can then be interpolated to different stellar parameters. Such grids have been computed, in 3D non-LTE, for: H \citep{Amarsi18a}, Li \citep{Sbordone10,Mott20,Wang21}, C \citep{Amarsi19b}, O \citep{Amarsi16a,Amarsi19b}, Mg \citep{Matsuno24}, Fe \citep{Amarsi22}, Na \citep{Canocchi24arxiv} and Ba (Steffen M. in prep., in the 1.5D non-LTE approximation). 
These grids have been applied to relatively small samples of the order tens to hundreds of stars \citep{Amarsi19b, Giribaldi21,Giribaldi23,Francois24,Matsuno24}. However, recently \cite{Wang24} determined 3D non-LTE Li abundances for hundreds of thousands of stars in GALAH DR4 \citep{Buder24}, opening the door to similar applications for other elements. Regarding \ion{Ca}{II}, multiple grids of 1D non-LTE synthetic spectra or abundance corrections have been computed \citep{Starkenburg10,Merle11,Spite12,Osorio22,Mashonkina23} but a dedicated study in 3D non-LTE across the stellar parameter range of metal-poor FGK-dwarfs is still missing.

In this work, we computed a grid of 3D non-LTE synthetic spectra of the CaT and \hk{} lines using the recently published \stagger{}-grid of 3D model atmospheres \citep{Rodriguez24} as described in \sect{Sec:method}. We investigated the 3D non-LTE effects on line strengths and shapes in \sect{Sec:Results} and present 3D non-LTE abundance and radial-velocity corrections for the CaT lines. Finally, we discuss the potential uses and implications of these corrections in \sect{Sec:discussion}.

\section{Method}\label{Sec:method}

\begin{figure*}
  \centering
   \resizebox{7in}{!}{\includegraphics{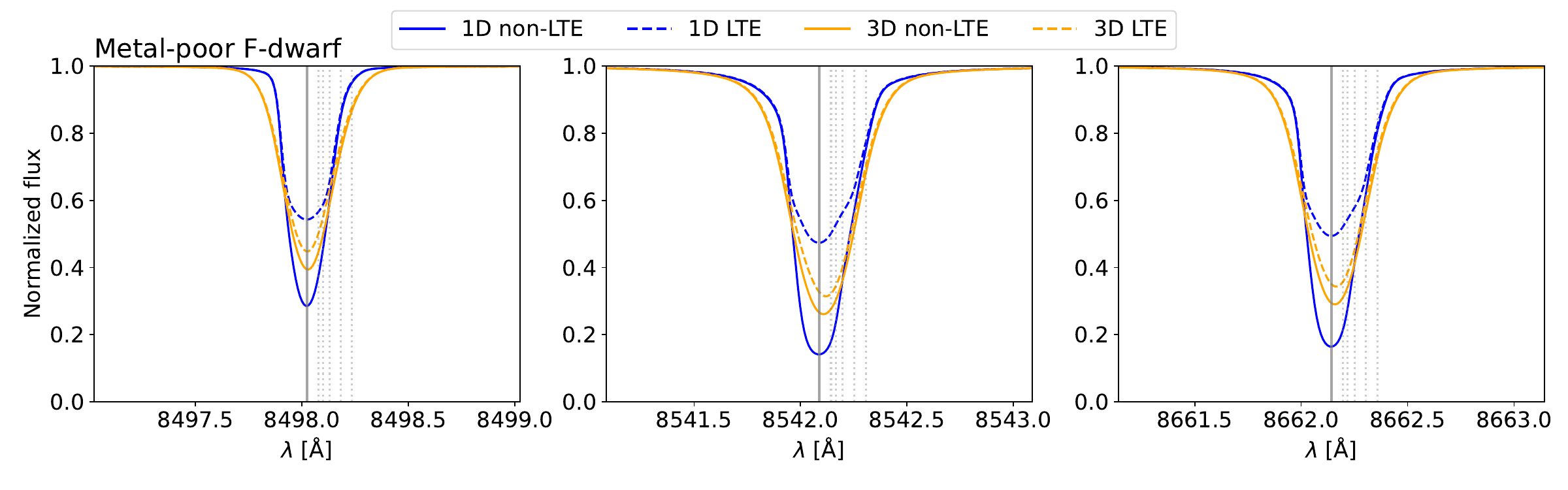}}
  \resizebox{7in}{!}{\includegraphics{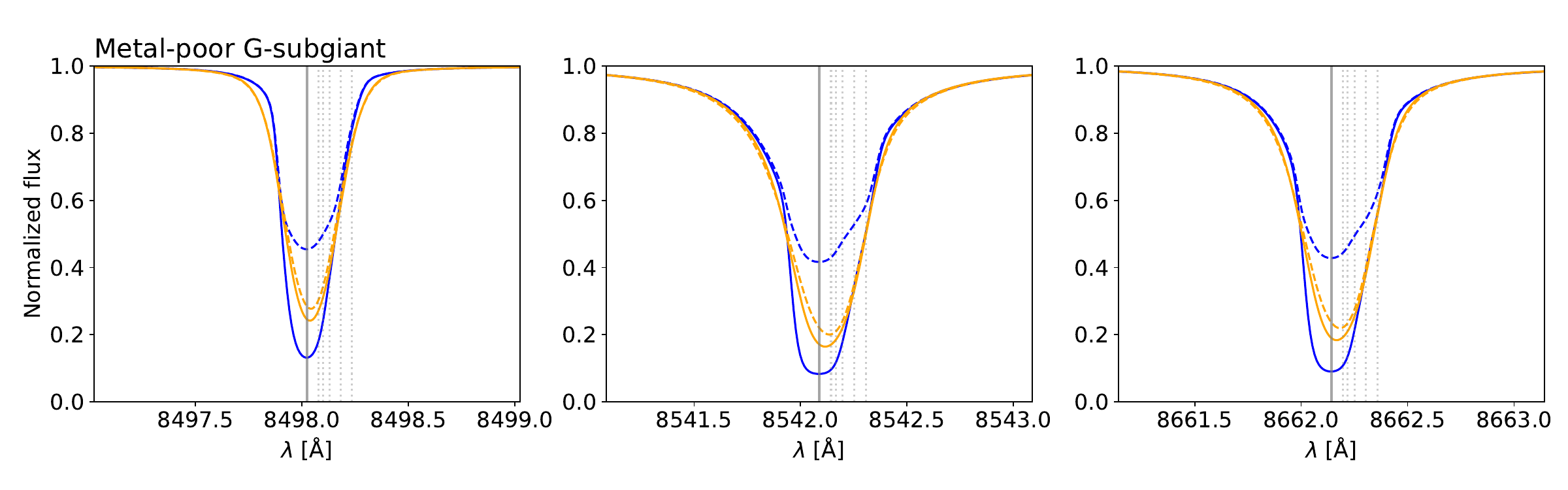}}
\caption[]{The CaT lines in 1/3D (non-)LTE. The line centre wavelengths are shown for the dominant isotope (solid grey) and other isotopes (dotted grey). Top panel: metal-poor dwarf with stellar parameters $T_\mathrm{eff}=6500$ K, $\log g=4.0$ and $\mathrm{[Fe/H]}=-3$. Bottom panel: metal-poor subgiant with stellar parameters $\teff=5500$ K, $\logg=3.0$ and $\feh=-3$. All spectra is computed at an abundance corresponding to $\mathrm{[Ca/Fe]}=+0.4$, microturbulence $\vmic=1.5~\kms$ (for the 1D synthesis) and no macroturbulent broadening. 
} 
\label{fig:CaT_all}
\end{figure*}

\subsection{Model atmospheres}\label{sec:modelatmos}

The 3D model atmospheres are taken from the updated, publicly available\footnote{\url{https://3dsim.oca.eu/fr/the-stagger-grid-2-0}}, \stagger{}-grid \citep{Rodriguez24}, computed using the \stagger{}-code \citep{Nordlund95,Magic13a,Collet18,Stein24}. 
The models are labelled by effective temperature $\teff$, surface gravity $\logg$, and iron abundance $\feh$ as a proxy for the overall chemical composition.  The composition is that of the Sun as given by \citet{Asplund09}, with metal abundances scaled by $\feh$. The models with $\feh\leq-1.0$ are alpha-enhanced with $\mathrm{[\alpha/Fe]}=+0.4$. 
An overview of the grid is shown in \fig{fig:gridstatus}. Note that the effective temperature of a \stagger{} model is an emergent property such that the grid is not perfectly regular in $\teff-\logg$ space.

The \stagger{}-grid models for different stars are set up such that they all contain approximately ten granules and span at least three convective turnovers. For the 3D non-LTE post-processing the models were downsampled following
the guidelines presented in
\cite{Rodriguez24}: by using fictitious Fe I lines synthesised in 3D LTE, they showed that the horizontal mesh can be downsampled from 240 cells to 80 cells without any significant impact on line shape and strength. In addition, as few as two snapshots (when carefully selected) are sufficient to obtain an equivalent width with maximum abundance error of $0.01~\dex$, whereas around 10 snapshots are sufficient to accurately model the line shape with a maximum flux error of 0.005. 
In this work we further explored the aforementioned post-processing methods, as described in  \app{sect:A_modelres}. We found that we could downsample the horizontal mesh to 48 cells in each direction, and reduce the number of snapshots used in our spectrum synthesis from 10 to 5 without introducing significant errors on line shape and strength, in the context of this work.

The 1D model atmospheres used in this work come from the public \marcs{} grid \citep{Gustafsson08}. The same grid configuration was used in \citet{Amarsi20b}.  In summary, for $\logg \geq 4.0$ (``dwarfs''), plane-parallel models with microturbulence $\vmic=1\,\kms$ were used,
whereas for $\logg \leq 3.5$ (``giants''), spherical models with $\vmic=2\,\kms$ and of solar mass were used.  All models are of the ``standard'' \marcs{} composition: namely, they are based on the solar composition of \cite{Grevesse07}, scaled by $\feh$. All of the 1D models used in the present study, being metal-poor with $\feh\leq-1$, are alpha-enhanced with $\mathrm{[\alpha/Fe]}=+0.4$, similar to the 3D models. We stress that in the spectral line synthesis on the 1D models (\sect{Sec:synthesis}) the microturbulence was not constrained to match that of the model atmosphere but was instead treated as a free parameter.

In the \marcs{} grid and with the above constraints on geometry, microturbulence, mass, and composition, two models are missing from the public repository with stellar parameters $\teff=4500,~5000~\mathrm{K}$, $\logg=2.5$ and $\feh=-4$.
These missing models were computed by interpolating the atmospheric structure between neighbouring \marcs{} models at fixed effective temperature and surface gravity, but varying metallicity\footnote{Interpolation was performed using the public routine developed by Masseron T., available at \url{https://marcs.oreme.org/software/}.}. Four models were not available in spherical geometry, with stellar parameters: $\teff=5000,~5000,~5500,~5500~\mathrm{K}$, $\logg=3.0,~3.5,~3.0,~3.5$ and $\feh=-4,~-3,~-4,~-3$. We include these four (sub)giant models in plane-parallel geometry with standard composition and microturbulence of $\vmic=2\,\kms$, consistent with the other (sub)giants. The impact of the chosen geometry on the temperature structure, at a surface gravity of $\logg \ge 3$, is less than $0.01~\%$ (see Figs. 13 and 15 of \citealt{Gustafsson08}).

Each \marcs{} model is set on a 1D mesh of 56 cells covering a range in optical depth of $-5\le\log\tau_\mathrm{ross}\le2$. Since the effective temperature of the 1D \marcs{} models are fixed, contrary to the 3D models, any direct comparison of spectral diagnostics that are computed with the 3D \stagger{} models and the 1D \marcs{} model will introduce some error due to the mismatch in effective temperature, that varies from model to model. The majority of 3D models have an emergent $\teff$ that lies within $25$ K of the 1D model, while 12 models deviate between $50$ and $100$ K. To make a consistent analysis, we interpolated the computed equivalent widths from the regularly spaced 1D grid to the irregular spaced 3D grid points. Specifically, we used cubic spline interpolation to interpolate the equivalent width between at least three neighbouring 1D models at varying effective temperature but equal metallicity, surface gravity, and microturbulence.

\subsection{Model atom}\label{Sec:modelatom}
The six-level model used for the 3D non-LTE iterations consists of five energy levels of singly-ionised calcium and the ground level of doubly-ionised calcium.  In the final synthesis the departure coefficients were redistributed onto a model atom with 52 levels for population conservation.

The energies were taken from \citet{Sugar85} via the National Institute of Standards and Technology Atomic Spectra Database (NIST ASD; \citealt{Ralchenko20}).  The six-level model includes the \ion{Ca}{II} \hk{} and triplet lines with oscillator strengths from \cite{Theodosiou1989}, as well as the forbidden lines coupling the $\mathrm{4d\,^{2}S^{o}}$ ground state to the $\mathrm{3d\,^{2}D^{o}}$ lower level of the CaT with oscillator strengths from \citet{Osterbrock51} via NIST ASD; these are illustrated in \fig{fig:term_diagram}. All five energy levels of singly-ionised calcium are coupled to the continuum via photo-ionisations with cross-sections taken from The Opacity Project (TOPBase; \citealt{Cunto93}).  Rate coefficients for electron collisional excitation were taken from The Iron Project (TIPBase; \citealt{Melendez07}), while for collisional ionisation the recipe from \citet{Allen1973} was adopted.  Rate coefficients for collisional excitation and charge exchange with neutral hydrogen were taken from \cite{Belyaev18}.  Pressure broadening by hydrogen collisions for the \ion{Ca}{II} \hk{} and CaT is based on ABO theory with the coefficients taken from \citet{Barklem98}.

Isotopic splitting was taken into account. The isotopic shifts were calculated from the Ritz wavelengths of the energy levels of the major isotopes \citep{MartenssonPendrill1992,Nortershauser1998}, and the isotopic fractions are those given in Table B1 of \citet{Asplund21}, which are based on the values measured in CI chondrites. \cite{Leenaarts14} showed that in order to accurately model the characteristic inverse-C shape of the line bisector of the solar \ion{Ca}{II} $8542~\AA$ line \citep{Uitenbroek06}, one needs to include isotopic splitting. In addition, the impact and presence of \ion{Ca}{II} isotopes has been thoroughly studied for chemical peculiar stars \citep{Castelli04,Cowley05,Cowley07,Ryabchikova08}.

Overall, the model is similar to the one used in the 3D non-LTE solar chromosphere study of \citet{Leenaarts14} and \citet{Bjorgen18}. The main difference relevant to this work on stellar photospheres is the inclusion of inelastic collisions with neutral hydrogen, which were neglected in these studies.  Here, data are taken from \cite{Belyaev18}. Hydrogen collisions grow in importance towards lower metallicities as the number of free electrons becomes smaller.

\subsection{Spectrum synthesis}\label{Sec:synthesis}

\begin{figure*}
  \centering
  \resizebox{7.0in}{!}{\includegraphics{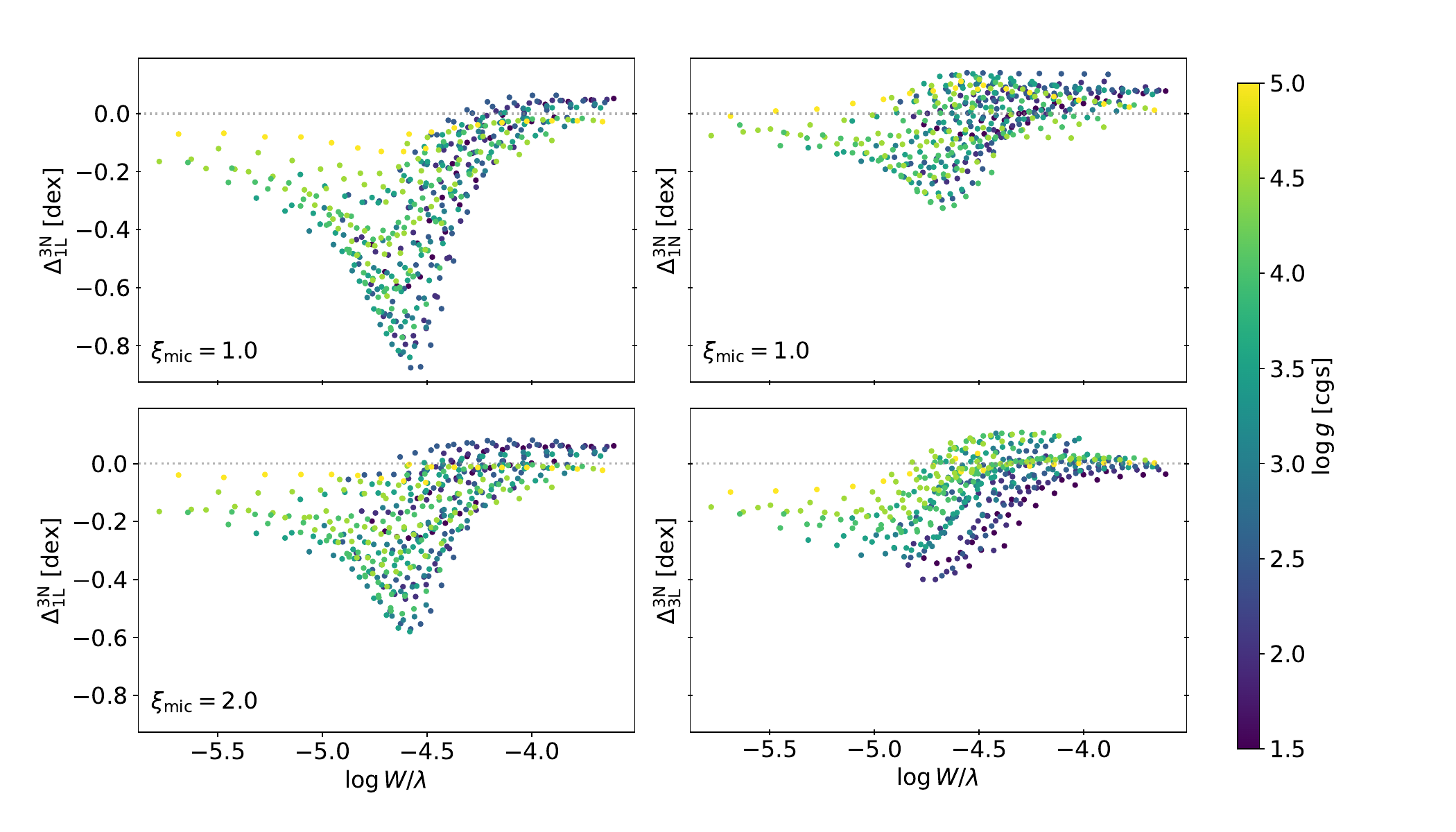}}
\caption[]{3D non-LTE abundance corrections as compared to 1D (non-)LTE and 3D LTE for the CaT $8497$ line at $\mathrm{[Ca/Fe]}=-0.6~\mathrm{to}~+1.4~\dex$ in steps of 0.25 dex, and $\vmic=1.0,~2.0\,\kms$.} 
\label{fig:ACacorr_8497_saturation}
\end{figure*}

\begin{figure*}
  \centering
  \resizebox{7.0in}{!}{\includegraphics{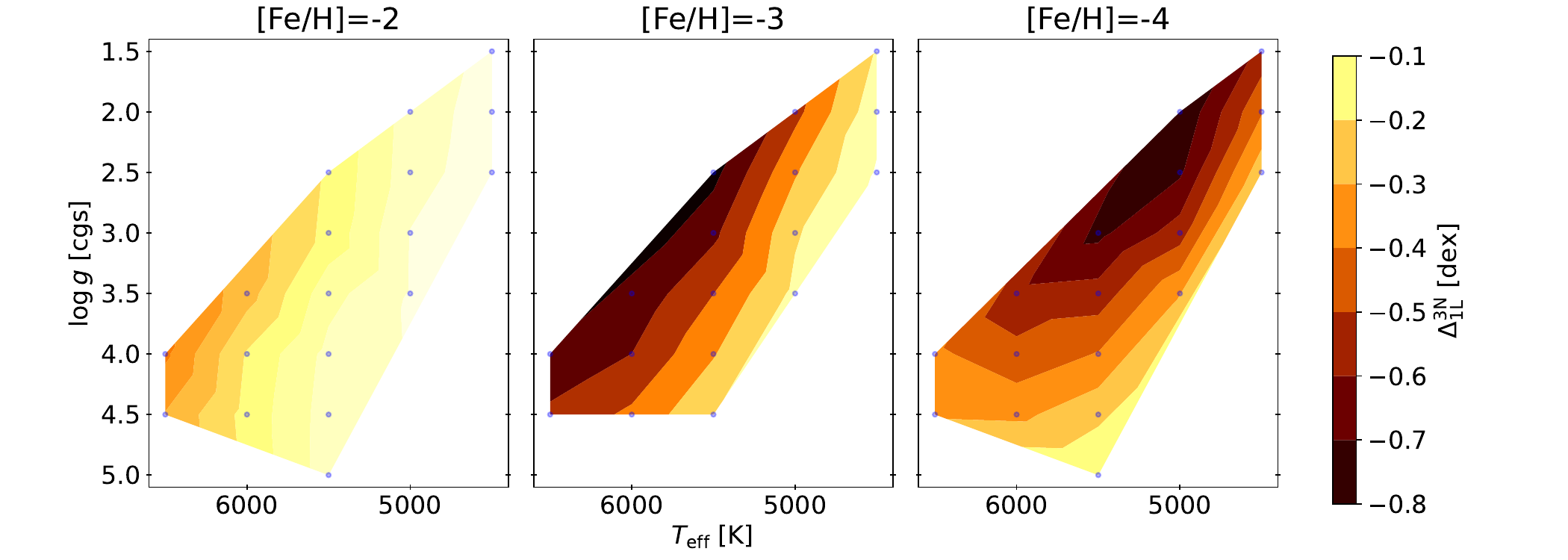}}
\caption[]{3D non-LTE abundance corrections as compared to 1D LTE for the CaT $8497$ line at $\mathrm{[Ca/Fe]}=+0.4~\dex$ and $\vmic=1.0\,\kms$. } 
\label{fig:HRD_dACa_8497_a100_CaFep040}
\end{figure*}

\begin{figure}
  \centering
  \resizebox{3.4in}{!}{\includegraphics{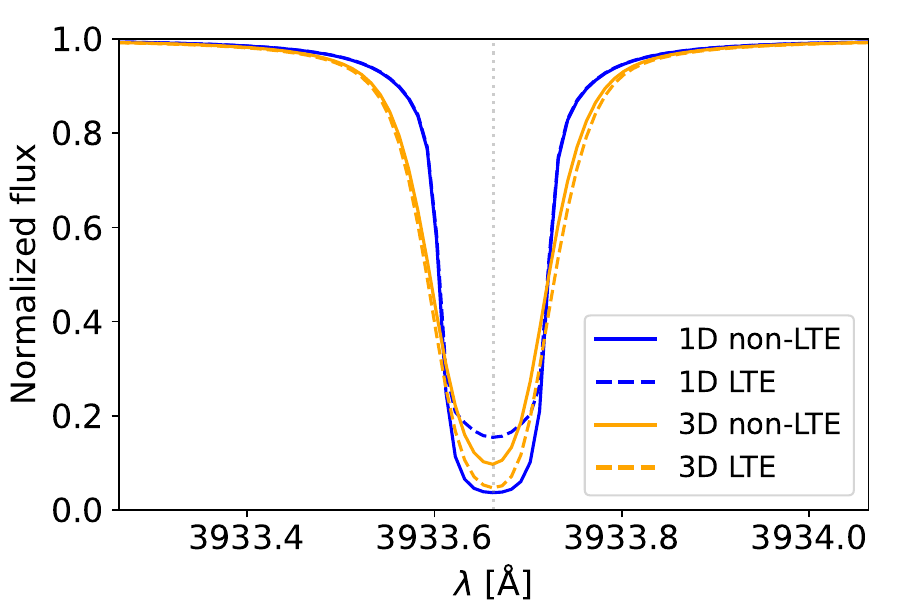}}
\caption[]{The \ion{Ca}{II} K line in 1/3D (non-)LTE for the metal-poor dwarf with $\teff=6500$ K, $\logg=4.5$, $\feh=-4$ and $\mathrm{[Ca/Fe]}=-0.6$. For the 1D synthesis, extra microturbulent broadening was used corresponding to $\vmic=1.0~\kms$ while no macroturbulent broadening was applied.} 
\label{fig:CaK_line_t65g45m40_CaFem060}
\end{figure}

\begin{figure*}
  \centering
  \resizebox{7.0in}{!}{\includegraphics{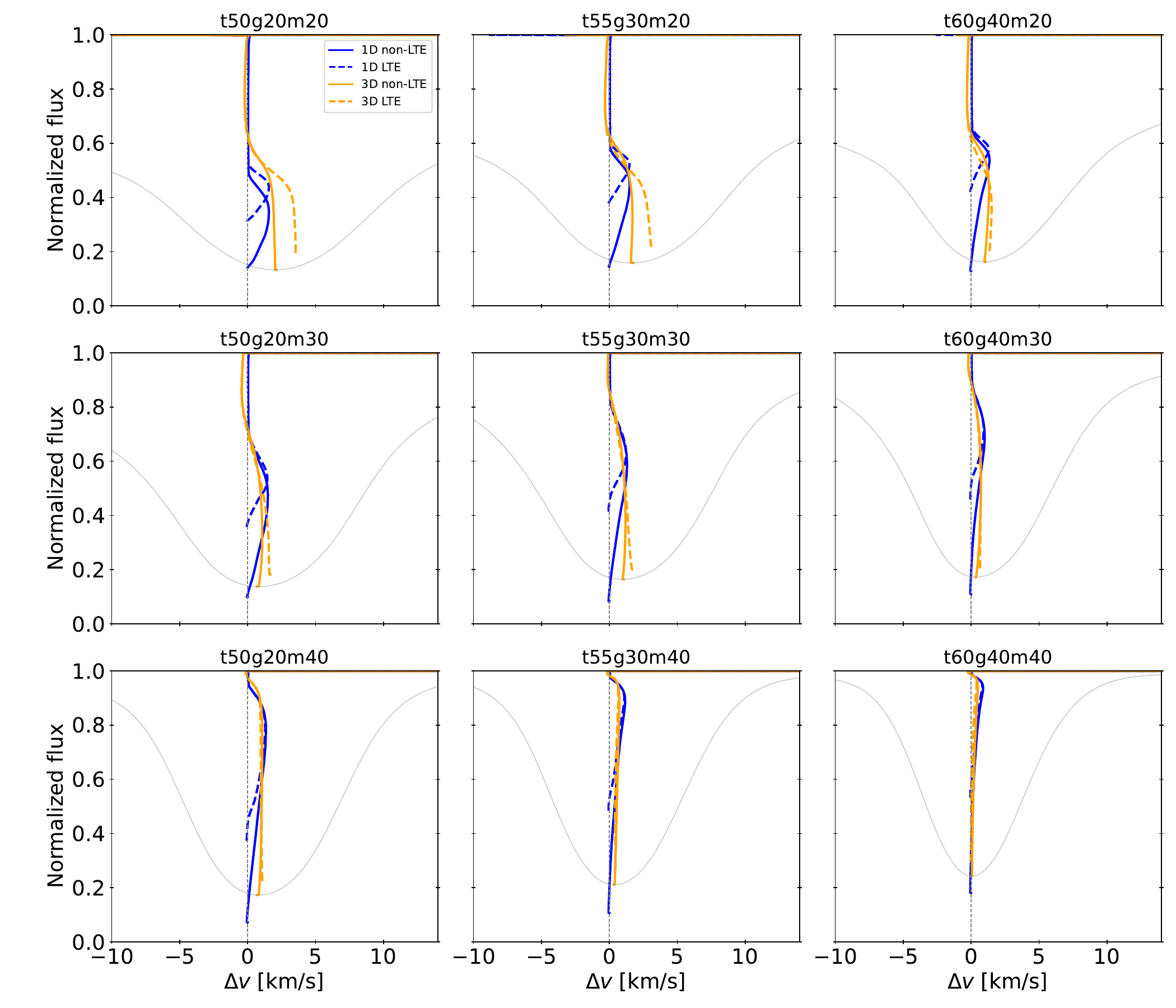}}
\caption[]{Overview of the CaT $8542~\AA$ bisector for three stellar models and three metallicities with a calcium abundance corresponding to $\mathrm{[Ca/Fe]}=+0.4~\dex$, computed in 1D using a microturbulence of $\vmic=1.5\,\kms$. The grey dotted line marks line centre and the 3D non-LTE line profile is shown in the background (solid grey). Macroturbulent broadening was not applied to the 1D spectra.} 
\label{fig:bisectors}.
\end{figure*}

\begin{figure*}
  \centering
  \resizebox{7.0in}{!}{\includegraphics{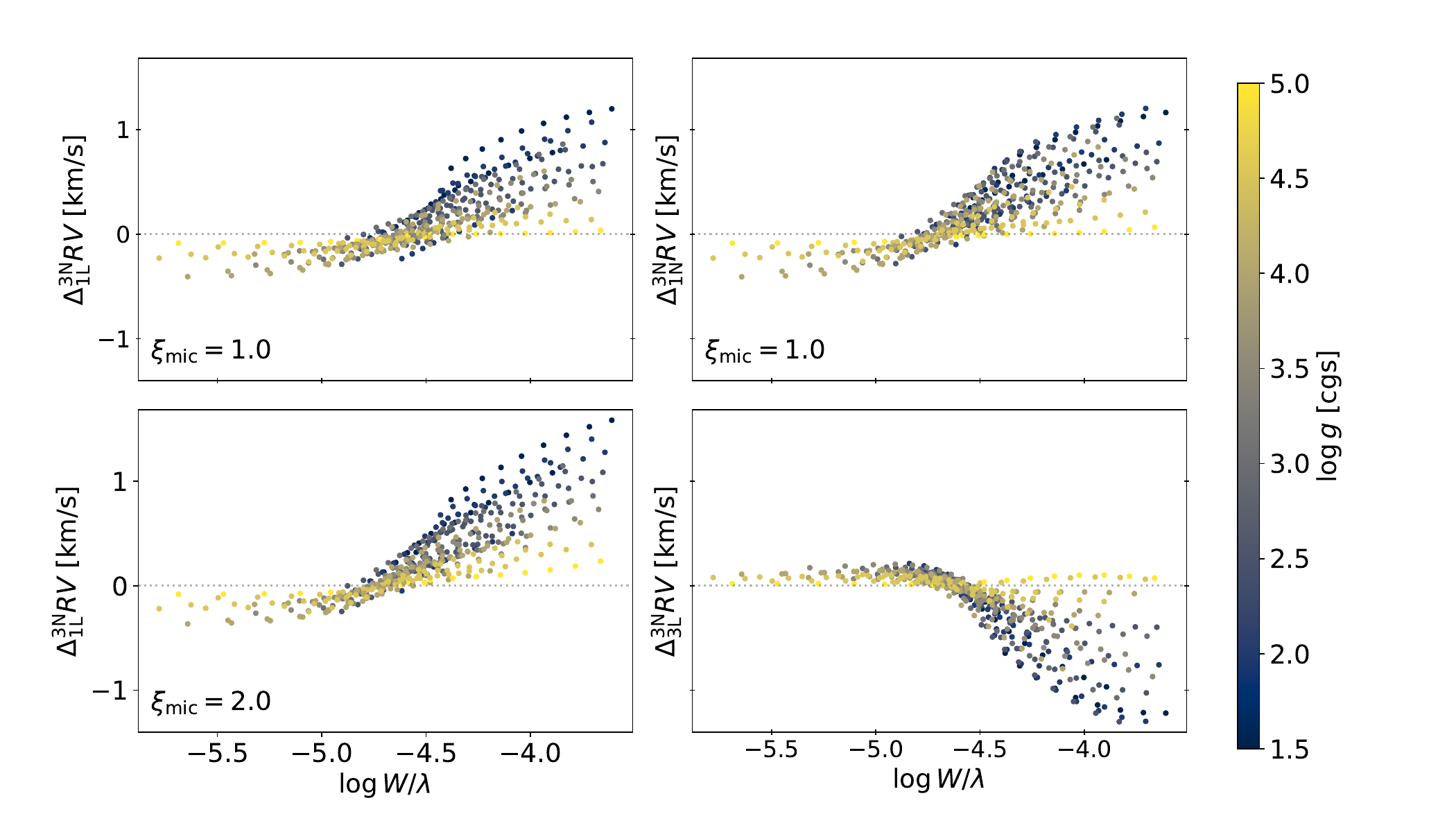}}
\caption[]{3D non-LTE radial velocity corrections as compared to 1D (non-)LTE and 3D LTE for the CaT $8497$ line at $\mathrm{[Ca/Fe]}=-0.6~\mathrm{to}~+1.4~\dex$ in steps of $0.25~\dex$, $\vmic=1.0,~2.0\,\kms$ and no additional macroturbulent broadening.} 
\label{fig:RVcorr_8497_saturation}
\end{figure*}

\begin{figure}
  \centering
  \resizebox{3.5in}{!}{\includegraphics{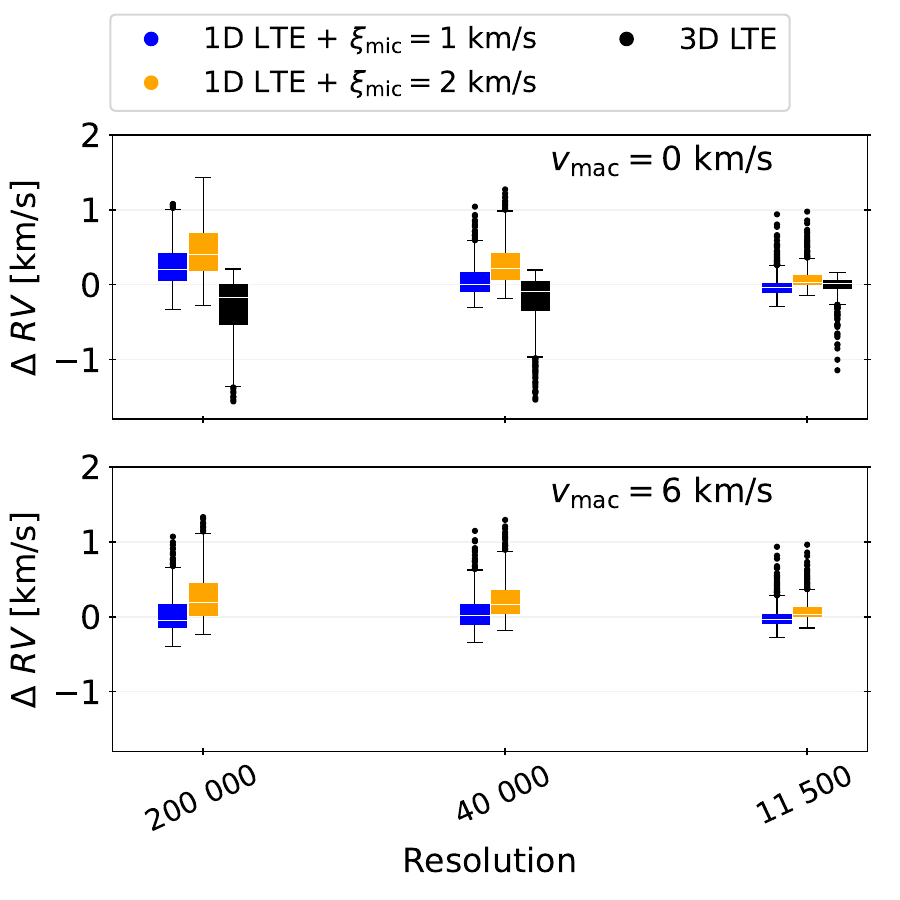}}
\caption[]{Compilation of all radial velocity corrections for the three CaT lines combined, all stellar parameters and abundances, and macroturbulence applied to the 1D profiles (row). Corrections to 1D LTE with $\vmic=1.0$ and $2.0\,\kms$ are shown as blue and orange boxplots, respectively, with 3D LTE shown in black.} 
\label{fig:RVcorr_boxplots}
\end{figure}

The synthetic spectra in this work were computed using the 3D non-LTE radiative transfer code \balder{} \citep{Amarsi18a}, which is based on \texttt{Multi3D} \citep{Botnen99,Leenaarts09b} with various updates in particular to its treatment of background opacities
\citep[e.g.][]{Amarsi16b}. In particular, the code used here includes a treatment of the dissolution of Rydberg hydrogen levels based on the HBOP module within the HLINOP 
package \citep[e.g.][]{Barklem16}.  The main effect of this is the smooth continuation of continuous opacity beyond the Balmer and Paschen breaks which has a small effect on the line profiles of 
the \ion{Ca}{II} \hk{} and CaT respectively.

\balder{} solves the restricted non-LTE radiative transfer using the trace-element approximation for the elements of choice.  
In our final results, we restricted ourselves to only compute the non-LTE level populations of \ion{Ca}{II}, under the assumption that changes in the background opacity due to non-LTE effects of other elements are negligible. We tested the impact of this assumption for a metal-poor giant, by simultaneously treating calcium and iron in 1D non-LTE, using the iron model atom from \citet{Amarsi22}, and taking care to use the same iron model atom when modelling it in LTE. We found that non-LTE effects of iron have a negligible impact on the line shape and strength of the CaT and \hk{} lines at the investigated metallicities (see \app{sect:A_multi_element}).
We note that although \cite{Osorio20} found that non-LTE effects on magnesium may affect the level populations of neutral calcium in the Sun via changes to the UV opacity, later studies found negligible inter-species effects \citep{Asplund21,Osorio22}. In any case, such effects would be smaller towards lower metallicities where hydrogen increasingly dominates the overall opacity.  

The non-LTE iterations for the 3D spectral synthesis were performed with a 8-point Lobatto quadrature over $\mu = \cos\theta$ (which includes the upwards and downwards vertical) and the 4-point trapezoidal integration over $\phi$ for inclined rays, amounting to 26 points on the unit sphere \citep{Amarsi17}. We investigated the impact of the chosen angle quadrature as compared to a reference case in  \app{sect:A_angledep}. The level populations were deemed to have converged once the relative maximum change of the emergent intensities between iterations is smaller than $0.1\%$. After convergence of the level populations, the final emergent spectrum was synthesised with a higher wavelength resolution in the CaT and \hk{} lines. 

For each 3D model atmosphere and each of its five snapshots, synthetic spectra of the CaT and \hk{} lines were computed for nine different calcium abundances between $-0.6 \leq \mathrm{[Ca/Fe]} \leq +1.4$ in steps of $0.25~\dex$. Over the full grid this corresponds to a range in calcium abundance of $1.70<A(\mathrm{Ca})<5.70$. The large range in calcium abundance was chosen to capture the increased scatter in $\mathrm{[Ca/Fe]}$ at low metallicities \citep{Francois20}. 

Spectrum synthesis on 1D model atmospheres requires extra free parameters to mimic the effects of convective motions in the stellar atmosphere: namely micro- and macroturbulence (and various mixing length parameters need to be considered when constructing the 1D model atmosphere itself). We computed 1D synthetic spectra with \texttt{Balder} for three values of microturbulence: $\vmic=1.0$, $1.5$, and $2.0\,\kms$. As we mentioned above (\sect{sec:modelatmos}), the microturbulence in the synthesis was not constrained to match the one used to construct the 1D \marcs{} model (namely $1.0\,\kms$ for dwarfs and $2.0\,\kms$ for giants). Macroturbulence does not affect equivalent widths and hence abundance corrections, but it does affect radial velocity corrections. As such, in \sect{sec:bisectors} and \sect{sec:RVcorrs} we applied macroturbulence to the 1D synthetic spectra by convolving the synthetic spectra with Gaussian kernels of specified full-widths at half-maxima. We also stress that no additional micro- or macroturbulence (or any other fudge parameter) was employed in the 3D non-LTE spectrum synthesis.

\section{Results}\label{Sec:Results}

\subsection{Departures from LTE in 1D and 3D}
Ca is mostly singly ionised in FGK-type stellar photospheres and can be treated as the majority species over \ion{Ca}{I}. Consequently, the \ion{Ca}{II} \hk{} resonance lines experience little to no non-LTE effects, except in the cores that form in the uppermost layers of our photosphere models.

The non-LTE effects on the CaT lines in 1D models have previously been discussed by \citet{Mashonkina07,Sitnova19,Neretina20}. We build upon their work by also considering the impact of hydrogen and electron collisions on the level populations. Specifically, we investigated the nature of the 1D non-LTE effects ourselves  by ``switching off'' certain radiative or collisional bound-bound transitions in the model atom, allowing us to quantify the impact of each transition on the level populations. By doing so, we confirm that, in 1D, the dominant non-LTE effect are photon losses in the CaT lines themselves, from the upper $4\mathrm{p}~^2\mathrm{P}$ level to the metastable $3\mathrm{d}~^2\mathrm{D}$ level, reducing the population of the upper level. Since the lower $3\mathrm{d}~^2\mathrm{D}$ level is metastable, we do not expect strong photon losses to the $4\mathrm{s}~^2\mathrm{S}$ ground state (the 730 nm bound-bound transition connecting the two levels is electric-dipole forbidden).  However, the $3\mathrm{d}~^2\mathrm{D}$ level is strongly coupled to the $4\mathrm{s}~^2\mathrm{S}$ ground state through hydrogen and electron collisions, leading to a $3\mathrm{d}~^2\mathrm{D}$ level population that is close to LTE. As such, the photon losses in the CaT lines lead to a sub-thermal source function compared to LTE, in turn leading to stronger lines as seen in \fig{fig:CaT_all}. In addition, since photon losses occur predominantly in higher layers, they mainly affect the line core. \citep{Mashonkina07,Amarsi16a}.

For the \ion{Ca}{II} \hk{} lines, the $4\mathrm{s}~^2\mathrm{S}$ ground state acts as an  electron reservoir and hence does not deviate from LTE. The upper $4\mathrm{p}~^2\mathrm{P}$ level is underpopulated relative to LTE as mentioned above. This results in a sub-thermal source function in the \hk{} lines and therefore stronger lines in non-LTE as seen in \fig{fig:CaK_line_t65g45m40_CaFem060}.

Metal-poor 3D models have steeper temperature gradients than the corresponding 1D models \citep[e.g.][]{Asplund99}. This leads to an increased UV photon pumping \citep[e.g.][]{Amarsi16b}. Here, there is extra pumping of the \hk{} lines in the 3D models. The consequence is that the 3D non-LTE population of the upper $4\mathrm{p}~^2\mathrm{P}$ level is brought closer to 3D LTE, as compared to 1D non-LTE versus 1D LTE. This is reflected in \fig{fig:CaT_all} where the difference in line profiles between 3D LTE and 3D non-LTE is less pronounced than the difference between 1D LTE and 1D non-LTE.

Lastly, it can be seen in \fig{fig:CaT_all} that the addition of non-LTE in 3D leads to a change in the line core shape as compared to 3D LTE which will be further explored in \sect{sec:bisectors} and \sect{sec:RVcorrs}. Whereas in 1D LTE and in the absence of any macrotubulent broadening, the lines are sufficiently weak that the different isotope contributions to the line profile are visible.

\subsection{3D non-LTE abundance corrections}\label{Sec:abundcorr}

Abundance corrections for each CaT line were obtained by matching equivalent widths on the curve of growth between two model spectra, at given stellar parameters ($\teff$, $\logg$, $\feh$, 1D $\vmic$, and reference calcium abundance $\lgeps{Ca}$).
For example, the 3D non-LTE versus 1D LTE abundance correction was computed as follows:
\begin{equation}
    \Delta^\mathrm{3N}_\mathrm{1L}=\log\epsilon^\mathrm{3N}(W_\lambda) - \log\epsilon^\mathrm{1L}(W_\lambda)~~,
\end{equation}

with $\log\epsilon(W_\lambda)$ the abundance at a given equivalent width for the 1D LTE spectra ($\mathrm{1L}$). Here, we used the equivalent width corresponding to the nine abundances computed in 1D LTE, which, at a fixed metallicity, range from $\mathrm{[Ca/Fe]}=-0.6~\mathrm{to}~+0.4~\dex$, to sample the full parameter space. In the following discussion, we define weak lines as having a reduced equivalent width smaller than $\log W/\lambda \lessapprox -5$, saturated lines $-5 \gtrapprox\log W/\lambda \lessapprox -4$ and strong lines $\log W/\lambda \gtrapprox -4$. 

\subsubsection*{Relation with line strength}

We visualise all abundance corrections for each model and abundance in relation to the reduced equivalent width of the 1D LTE spectra, colour coded with the surface gravity of the model in \fig{fig:ACacorr_8497_saturation}. This allows us to investigate the relation between abundance correction with line strength and saturation. 

The main result is that the largest abundance corrections occur when the CaT lines are close to saturation, which happens at a reduced equivalent width of approximately $\log W/\lambda \approx -4.8$. Since the CaT lines saturate at different abundances depending on the assumption, 1/3D (non-)LTE, the largest difference between the curve of growths is found close to saturation. This relation between abundance corrections and saturation has been studied for other elements such as \ion{Fe}{I} \citep{Amarsi22}, \ion{K}{I} \citep{Reggiani19}, \ion{Na}{I} \citep{Lind11}, and \ion{Ca}{I} and \ion{O}{I} \citep{Amarsi19b}. Since the value of microturbulence in 1D dictates when lines saturate on the curve of growth, this reflects on the size of the resulting $\Delta^\mathrm{3N}_\mathrm{1L}$ abundance corrections. For both values of microturbulence $\vmic=1.0-2.0\,\kms$, the CaT lines saturate first in 1D LTE compared to 3D non-LTE, resulting in negative abundance corrections down to $\Delta^\mathrm{3N}_\mathrm{1L}>-0.95~\dex$. The corrections for weak lines range between $\Delta^\mathrm{3N}_\mathrm{1L}=0~\mathrm{to}~-0.5~\dex$. For increasing values of microturbulence, the 1D LTE lines saturates at higher abundances more similar to 3D non-LTE, leading to smaller abundance corrections. 

In the same \fig{fig:ACacorr_8497_saturation}, we investigated corrections relative to 1D non-LTE and 3D LTE. First, the relation between the abundance corrections relative to 1D non-LTE $\Delta^\mathrm{3N}_\mathrm{1N}$ and microturbulence behaves differently to the 1D LTE case. For a microturbulence of $\vmic=1.0\,\kms$, the corrections $\Delta^\mathrm{3N}_\mathrm{1N}$ are centred around zero with a stronger negative peak near saturation. For larger values of microturbulence $\vmic=2.0\,\kms$, the 3D non-LTE lines saturate first, leading to positive abundance corrections. Overall, the abundance corrections relative to 1D non-LTE $\Delta^\mathrm{3N}_\mathrm{1N}$ are closer to zero than the corrections relative to 1D LTE.

Lastly, a similar behaviour is observed for the corrections relative to 3D LTE $\Delta^\mathrm{3N}_\mathrm{3L}$. These corrections become negative near saturation and subsequently, for the weak lines, reach a similar magnitude as the $\Delta^\mathrm{3N}_\mathrm{1L}$ corrections.

\subsubsection*{Relation with stellar parameters}
In \fig{fig:HRD_dACa_8497_a100_CaFep040}, we visualize how the $\Delta^\mathrm{3N}_\mathrm{1L}$ abundance corrections change across the HR-diagram for a fixed metallicity and $\cafe$-ratio. Firstly, since the CaT lines grow weaker with increasing effective temperature, we observe the aforementioned correlation between line strength and abundance corrections clearly in the middle panel ($\feh=-3$). The CaT lines change from strong to saturated with increasing $\teff$, leading to more negative abundance corrections $\Delta^\mathrm{3N}_\mathrm{1L}$. In addition, we observe that at a fixed effective temperature, the abundance corrections grow more negative with decreasing surface gravity. This effect is more pronounced when the CaT lines are saturated.

\subsubsection*{The \ion{Ca}{II} \hk{} lines}

Unlike the CaT lines, the \hk{} lines stay saturated down to the lowest metallicity and abundance explored in this work. The weakest, but still saturated, \hk{} lines are found for the F-dwarfs at a metallicity of $\feh=-4$ and $\cafe=-0.6$ with a minimum line strength of $\log W/\lambda\approx-4.55$. Specifically, we visualize the weakest \ion{Ca}{II} K line in \fig{fig:CaK_line_t65g45m40_CaFem060}. From this Figure it is clear that non-LTE only affects the line core, both in 1D and 3D. The addition of 3D tends to strongly `desaturate' the line resulting in a weaker line core but wider wings. Similar to the CaT lines, the abundance corrections are strongly related to the difference in line saturation between 3D non-LTE and the other cases. Specifically, we find that the abundance corrections for the \ion{Ca}{II} K line relative to 1D LTE, with $\vmic=-1~\kms$, reach down to $\Delta^\mathrm{3N}_\mathrm{1L}\approx-0.4~\dex$ for the `weakest' lines in our grid. Contrary, in the case with $\vmic=-2~\kms$, the corrections for \ion{Ca}{II} K are positive and reach at most $\Delta^\mathrm{3N}_\mathrm{1L}\approx0.05~\dex$. Regarding corrections relative to 3D LTE, we find that these are positive and reach up to $\Delta^\mathrm{3N}_\mathrm{3L}\approx0.17~\dex$.

Due to the limited extent of our grid, we are unable to fully explore the trend of increasing abundance corrections with line saturation. The aforementioned corrections are expected to keep increasing/decreasing with decreasing line strength until the \hk{} lines reach the weak part of the curve of growth.

\subsection{Bisectors in 3D LTE and 3D non-LTE}\label{sec:bisectors}

The line bisector is a strong diagnostic that reflects the interplay between line formation and velocity fields in the stellar atmosphere. Weak lines that form close to the optical surface show a blueshifted bisector that is characteristic of  convective motions at the stellar surface.  \citep{Dravins1981,Dravins1999, Gray09}. 

\fig{fig:bisectors} shows an overview of the change in bisector shape for the $8542~\AA$ line in nine model atmospheres. These models were chosen as they sample the metal-poor stellar evolution tracks shown in \fig{fig:gridstatus}. In the 1D models the bisector shows an inverse C-shape, similar to what is observed in the Sun in the solar chromosphere \citep{Uitenbroek06}. Opposite to the regular, blueshifted, C-shape of photospheric spectral line bisectors \citep{Dravins21}. The formation of the inverse C-shape is purely a result of including the calcium isotopes in the spectrum synthesis \citep{Leenaarts14}. 

In 3D (non-)LTE, the presence of velocity fields results in an overall asymmetric line profile that is redshifted as compared to the 1D bisectors. We find that the photospheric velocity fields tend to flatten the inverse C-shape. In addition, there is a noticeable redshift of the line core which is more pronounced in the stronger lines; compare top panels to the bottom panels. This is in contrast with the commonly observed convective blueshift of photospheric lines \citep{Dravins21} and can be attributed to the presence of the so-called `reverse granulation layer' that forms above the optical surface \citep{Bigot08,AllendePrieto13}. Non-vertical radiative heating of intergranular downflows by the surrounding hot granules leads to a reversal of the granulation pattern in the layers above the photosphere, such that downflows are now hotter than upflows \citep{Rutten04,Leenaarts05,Cheung07}. Only the upper wing of the CaT lines, close to the continuum, forms in the photosphere which reveals itself as a blueshift of the bisector. The remaining, stronger, parts of the line form higher up in this layer of reversed granulation, leading to an overall redshift of the line bisector. 

Interestingly, in the 3D models the non-LTE effects lead to an additional shift of the line core. This is primarily due to the granular nature of non-LTE effects across the stellar surface. We note that there are also some small differences between the 1D LTE and 1D non-LTE models as well, which is possible because of the isotopic components breaking the symmetry of the line. In the 3D models this shift can be as large as $1~\mathrm{km}~\mathrm{s}^{-1}$ for the models with stronger CaT lines. This discrepancy in the line core between 3D LTE and 3D non-LTE could be of consequence when estimating radial velocities by simply selecting the Doppler shift of the line core. A more thorough estimation of the radial velocity correction between 3D non-LTE and 1D (non-)LTE and 3D LTE is performed in the next section.

\subsection{Radial velocity corrections}\label{sec:RVcorrs}

Each spectral line was cross-correlated between 3D non-LTE, acting as the reference ``observed'' spectrum, and 1D (non-)LTE and 3D LTE, which are the template spectra. The cross-correlation was performed at a reference high-resolution of $R=200~000$, intermediate resolution of $R=40~000$ \citep[similar to UVES,][]{Dekker00} and that of the Gaia radial velocity spectrometer $R\approx11~500$ \citep[Gaia RVS]{Katz23} \citep[also similar to the 2dF/AAOmega spectrograph used by the southern stellar stream spectroscopic survey (S5) with $R=10~000$,][]{Li19}. The step in radial velocity is $0.03\,\kms$, in a range of $\pm150\,\kms$ around the theoretical line centre \citep{Chiavassa18}. In addition, extra macroturbulent broadening corresponding to $0$, $2$, $4$, or $6\,\kms$ was applied to the 1D spectra to simulate the Gaussian broadening of spectral lines due to convective velocities. These values were chosen to be representative of the root-mean-square velocities at a depth of $\log\tau_\mathrm{Ross}=-2$ as seen in the \stagger{} models and are consistent with 3D model computations of the \texttt{CO5BOLD}-group \citep{Steffen13}.

In practice, computing the cross-correlation function ($\ccf$) at a specific wavelength consists of shifting the template spectrum of one single CaT line by $n$ pixels, corresponding to a radial velocity shift of $n\times\Delta v$: 
\begin{equation}
    \Delta v = c*\frac{\lambda_{i+1} -\lambda_i}{\lambda_0}~~.
\end{equation}
The value of the $\ccf$ at a certain radial velocity shift is then \citep{AllendePrieto13}: 
\begin{equation}
    \ccf(n) = \sum_{k=1}^N S_{k+n-N/2}\cdot O_k~~,
\end{equation}
with $O$ the reference 3D non-LTE spectrum and $S$ the template spectrum. The radial velocity shift between the two flux arrays is then equal to the peak of the $\ccf$ which can be determined using a variety of methods. Here we opted to fit the top three pixels with a quadratic function, as in \cite{AllendePrieto13}, to obtain the final value of the radial velocity.

\subsubsection*{Relation with line strength and stellar parameters}
Similar to the abundance corrections, we visualise the radial velocity corrections for the $8497~\AA$ line as a function of reduced equivalent width in \fig{fig:RVcorr_8497_saturation}. First, looking at the radial velocity corrections between 3D non-LTE and 1D LTE $\Delta^\mathrm{3N}_\mathrm{1L} RV$. we find that the $\Delta^\mathrm{3N}_\mathrm{1L} RV$ corrections becomes more positive with increasing line strength. Moreover, the corrections transitions to negative values when the line becomes weak. Similar behaviour is observed for the $\Delta^\mathrm{3N}_\mathrm{1N} RV$ corrections while we see the opposite for the corrections relative to 3D LTE $\Delta^\mathrm{3N}_\mathrm{3L} RV$. In all cases, the size of the radial velocity corrections increase with decreasing surface gravity, independent of the other stellar parameters, microturbulence and calcium abundance.

Nowadays, 3D LTE template spectra are used to compute radial velocities instead of 1D LTE spectra \citep{AllendePrieto13,Chiavassa18,Dravins21}. In the case of giants, the $\Delta^\mathrm{3N}_\mathrm{3L} RV$ corrections can become increasingly negative with increasing line strength, up to $\Delta^\mathrm{3N}_\mathrm{3L} RV<-1.3~\kms$. On the other side, for weak lines, all corrections diminish down to $0<\Delta^\mathrm{3N}_\mathrm{3L} RV<+0.2~\kms$, irrespective of surface gravity. 

Overall, the corrections are on the order of the theoretical gravitational redshifts. These redshifts are of size $+0.7-0.8\,\kms$ for solar mass dwarfs with $\logg \sim 4.5$; the Sun has a empirically derived gravitational redshift of $+0.638\pm0.006\,\kms$ \citep[][and theoretical prediction $+0.6331\,\kms$]{GonzalezHernandez20}. While for giants they amount to $0.02-0.03\,\kms$ at $\logg \sim 1.5$ \citep{AllendePrieto13}.  

\subsubsection*{Relation with spectral resolution and macroturbulence}

The impact of macroturbulent velocity in the 1D models and resolution is shown in \fig{fig:RVcorr_boxplots}. In this Figure we compiled all corrections, for all stellar parameters and abundances, in a single boxplot. The box is centred around the median and contains $50\%$ of all data points. Each whisker adjacent to the box contains $24.65\%$ of all data points, when the data is normally distributed. In the following discussing we can assume that approximately $99\%$ of all data points is located in the box and whiskers. Outliers are then visualised by individual black points. 

As such, it is possible to visualise the bulk change in radial velocity corrections between different values of macroturbulence and resolution.  Including macroturbulent broadening in the 1D spectra results in lower radial velocity corrections. Moreover, macroturbulent broadening only significantly impacts the radial velocity corrections in the spectra at high resolution $R=200~000$. Values for macroturbulence greater than $v_\mathrm{mac}\ge4~\kms$ lower the 1D LTE radial velocity corrections, with $\vmic=1~\kms$, by $\approx 0-0.4~\kms$. A slightly smaller impact is seen for the 1D LTE corrections with $\vmic=2~\kms$, where the difference is $\lessapprox 0.35~\kms$. In general, decreasing the resolution to $R=40~000$ lowers the impact of macroturbulence to a maximum difference of $\approx0.1~\kms$. Decreasing the spectral resolving power from $R=200~000$ to $R=40~000$ reduces the range of corrections by half. Decreasing the spectral resolution further, down to the value representative for Gaia RVS, results in radial velocity corrections smaller than $\lesssim\pm0.25~\kms$, barring some outliers.

%--------------------------------------------------------------------
\section{Discussion}\label{Sec:discussion}

\subsection{Calcium abundances in very metal-poor stars ($\feh\lesssim-2$)}

Calcium is typically the only element present in EMP stars which has observable spectral lines in two ionisation stages. Furthermore, the CaT lines are expected to become unsaturated and form fully in the photospheres of ultra metal-poor stars (UMP: $\mathrm{[Fe/H]}\leq-4$; \citealt{Beers05}). At even lower metallicities, the \hk{} lines are used for abundance analysis instead of the CaT lines, together with the strong \ion{Ca}{I} resonance line at $4226~\AA$. As such, recovering equal elemental abundances for both \ion{Ca}{I} and \ion{Ca}{II} (``ionisation balance''), can be used to constrain stellar parameters \citep{Sitnova19} and the calcium abundance itself. For example, \cite{Norris07}, \cite{Mashonkina07}, \cite{Korn09} and \cite{Caffau12} utilised ionisation balance of calcium to constrain the stellar surface gravity of ultra metal-poor stars; while for the most metal poor star known to date (SDSS$\,$J102915.14+172927.9; the Caffau star), \citet{Lagae23} demonstrated its use as a consistency check on its surface gravity and its classification as a dwarf, as indicated by Gaia DR2 parallaxes.

From the grid of CaT lines we find that 1D LTE abundances are typically overestimated for metal-poor stars.  The most severe 3D non-LTE versus 1D LTE abundance corrections ($\Delta^\mathrm{3N}_\mathrm{1L}$) occur for the UMP giants. 
They are typically negative and can reach $-0.8\,\dex$ (i.e. the lines are stronger in 3D non-LTE compared to 1D LTE).
This is important because these are typically the brightest metal-poor stars observed. On the other hand, even for the UMP dwarfs we observe significant corrections relative to 1D LTE of the order $-0.2~\dex$. When the lines are in the weak part of the curve of growth, the closest approximation to 3D non-LTE tends to be 1D non-LTE, rather than 1D LTE or 3D LTE; but for saturated and strong lines, 3D LTE appears closest to 3D non-LTE.  These conclusions are in line with previous work by \cite{Lagae23} who found that 1D non-LTE and 3D non-LTE modelling of the CaT lines gave consistent calcium abundances within the uncertainties.

The cores of the strongest lines in our grid form at the uppermost layers of the model atmospheres, and line formation at those heights has to be treated with caution \citep{Osorio22}. To examine the sensitivity of the vertical extent on the synthetic spectra and resulting abundance corrections, we linearly extrapolated the \marcs{} models from $\log\tau_\mathrm{ross}=-5$ to $\log\tau_\mathrm{ross}\approx-7$, and computed a new grid of 1D (non-)LTE synthetic spectra with $\xi_\mathrm{mic}=1.5~\kms$.  Extrapolating to lower temperatures allows the line cores to grow deeper.  This reflects an extreme case, given that the temperature profile may become more shallow and eventually start to increase at these layers with the onset of the chromosphere.  In any case, we found that weak lines with $W_\lambda/\lambda\lessapprox-4.9$ are not sensitive to this test.  On the other hand, saturated lines of strength $-4.0\lessapprox W_\lambda/\lambda\lessapprox-4.9$ are sensitive to the vertical extent. Keeping the 3D non-LTE synthesis fixed, in the worse case the $\Delta^\mathrm{3N}_\mathrm{1L}$ abundance corrections change by $0.07\,\dex$, relative to a correction that reaches $0.9~\dex$. On quick inspection, the non-LTE results here are slightly more affected by the extension of the 1D model atmospheres, compared to what was reported in \cite{Osorio22}, at least for HD 122563. There could be many reasons for this including the choice of model atmosphere or the method of extrapolation, but investigating this is beyond the scope of the present work. Besides, we stress that 3D-1D abundance corrections for saturated lines are also always dependent on the choice of microturbulence in 1D, as visualised in \fig{fig:ACacorr_8497_saturation}. A change in microturbulence of $\pm0.25~\kms$ leads to a change of the $\Delta^\mathrm{3N}_\mathrm{1L}$ abundance corrections between $\pm0.04$ and $\pm0.07~\dex$.

\subsection{Calcium abundances in hyper metal-poor stars ($\feh\lesssim-5$)}

Obtaining accurate calcium abundances for stars of metallicities even lower than our grid is important as they are directly linked with the properties of Pop III stars and their nucleosynthesis. In present day stars, calcium is formed as part of the alpha process during He-burning. By contrast, since the first stars in the universe did not contain metals, their CNO cycle occurred at higher temperatures, potentially leading to enhanced calcium production through a process called `hot CNO breakout' (see \citealt{Wiescher1999}, \citealt{Clarkson21}, and references therein). Extensive work has been made to characterise the nuclear burning condition in Pop III stars from a theoretical standpoint \citep{deBoer21,Zhang22,Chen24}. 

At such low metallicities, calcium abundance determinations typically rely on the \ion{Ca}{II} K line, either in 1D LTE \citep{Aoki06b,Keller14,Frebel19,Aguado19} or 3D LTE \citep{Collet06,Frebel08,Keller14}. However, due to the limit of our grid we do not fully cover the curve of growth for these lines. Nevertheless, it is expected that the abundance corrections for these lines follow similar behaviour as the CaT lines: namely, larger corrections close to saturation. 

Specifically, \cite{Clarkson21} describes the need for a factor ten increase in nuclear rate to reproduce current observational calcium abundances in the most iron-deficient star \citep{Keller14}: namely, $A(\mathrm{Ca})=-0.60$, based on the 
3D non-LTE analysis of the \ion{Ca}{II} K line by \citet{Nordlander17a}. This corresponds to a $+0.44\,\dex$ correction relative to 1D LTE, with microturbulence $2\,\kms$. 
Although our current grid does not reach such low calcium abundances (minimum $A(\mathrm{Ca})=1.70$), this correction appears qualitatively consistent with what we would get if we extrapolate the trends with stellar parameters and abundances.  In the next paper we will expand our grid to lower calcium abundances to tackle the observed discrepancy between observations and theoretical predictions from Pop III nucleosynthesis.

\subsection{Radial velocities with Gaia RVS and other surveys}
Due to the strength of the CaT lines, even at low metallicities, they have been used to determine radial velocities such as in Gaia RVS \citep{Katz23}. Gaia RVS \citep{Katz23} and similar large-scale surveys such as the Southern Stellar Stream Spectroscopic Survey \citep[S5;][$R \sim 10~000$]{Li19} and the Radial Velocity Experiment \citep[RAVE;][$R \sim 7~500$]{Steinmetz20} aim for a radial velocity accuracy of $1.3-1.4~\kms$. The Gaia RVS is centred around the CaT region with an average resolving power of $11~500$ \citep{Katz04,Cropper18,Katz23}. The strong CaT lines can dominate the overall radial velocity signal \citep{AllendePrieto13,Chiavassa18} and hence a proper modelling of their shapes and strengths is crucial.

At the resolving power of Gaia RVS, the 3D non-LTE versus 1D LTE radial velocity corrections are in the range $-0.25\lesssim\Delta^\mathrm{3N}_{1L} RV\lesssim +0.25~\kms$ with a small sample of outliers reaching corrections as large as $\Delta^\mathrm{3N}_{1L} RV\sim\pm1.0~\kms$. 
(The 3D non-LTE versus 3D LTE radial velocity corrections are in a similar range, at least at this resolution.)
The outliers correspond to strong lines in cool giants at metallicity $\mathrm{[Fe/H]=-2}$ and need to be treated with caution. For these cases, our assumption that the line core forms fully in the stellar photosphere may break down and hence skew the shape of the line core. The properties of chromospheres in stars other than the Sun, and line formation therein, is still poorly understood, much less the computation of such atmospheres in 3D (see the discussion in \citealt{Wedemeyer17}). 

For high resolution observations of single stars, such as with the MAROON-X spectrograph \citep[$R \sim 85~000$]{Seifahrt20} which aims for an accuracy of $\approx1~\ms$, our corrections can be significant relative to 1D LTE: $-0.4\lesssim \Delta^\mathrm{3N}_\mathrm{1L}RV\lesssim +1.2~\kms$, and relative to 3D LTE: $-1.4\lesssim \Delta^\mathrm{3N}_\mathrm{3L}RV\lesssim +0.25~\kms$.
Following our analysis in \sect{sec:RVcorrs}, we find that at medium and high resolution, the impact of 1D macroturbulence is small relative to these corrections: the change in $\Delta^\mathrm{3N}_\mathrm{1L}RV$ is at most $\pm0.2~\kms$. At lower resolution the relative change in the radial velocity correction is negligible.

\section{Conclusions}\label{Sec:conlusion}

In this work we investigated the impact of \ion{Ca}{II} line formation in 3D non-LTE on the determination of abundances and radial velocities for metal-poor FGK-type stars. Grids of the \ion{Ca}{II} \hk{} and CaT line profiles were computed in 1/3D (non-)LTE for 53 metal-poor stellar atmospheres, for a broad range of calcium abundances and 1D microturbulence parameters. These grids were then used to calculate abundance and radial velocity corrections. Specifically, the grids of synthetic spectra, containing both the CaT and \hk{} lines, and the corrections only for the CaT lines will be made public.

The line profiles and abundance corrections can then be readily applied to individual stars or implemented in large surveys similar to what has been done in GALAH DR4 \citep{Wang24,Buder24}.  
Given the large 3D non-LTE effects on the commonly-used \ion{Ca}{I} resonance line at $4226~\AA$, these data will allow future studies to more accurately employ the CaT lines to derive calcium abundances in UMP stars.  The next paper in this series will expand this grid to even lower calcium abundances.

Besides abundance determinations, the CaT lines are used to derive radial velocities. We conclude that the correction in radial velocity due 3D non-LTE line formation, as compared to simpler assumptions, can be significant (of the same order as the stellar gravitational redshifts) when dealing with medium or high resolution spectra. For low resolution spectra, such as in Gaia RVS, the radial velocity corrections fall below the proposed accuracy of these instruments but could in some cases contribute significantly to the error budget.
Future studies could incorporate the 3D non-LTE CaT line profiles presented in this work with 1D non-LTE spectra of iron lines in the Gaia RVS range, for more accurate stellar characterisation.

\begin{acknowledgements}
We thank the referee for their constructive feedback. CL and KL acknowledge funds from the European Research Council (ERC) under the European Union’s Horizon 2020 research and innovation programme (Grant agreement No. 852977). CL also acknowledges funds from the UKRI Future Leaders Fellowship (MR/S035214/1). KL also acknowledges funds from the Knut and Alice Wallenberg foundation. AMA gratefully acknowledges support from the Swedish Research Council (VR 2020-03940) and from the Crafoord Foundation via the Royal Swedish Academy of Sciences (CR 2024-0015). The computations were enabled by resources provided by the National Academic Infrastructure for Supercomputing in Sweden (NAISS), partially funded by the Swedish Research Council through grant agreement no. 2022-06725, at the PDC Center for High Performance Computing, KTH Royal Institute of Technology (project numbers PDC-BUS-2022-4, NAISS 2023/1-15 and NAISS 2024/1-14). In addition, computations are enable by the Centre for Scientific Computing, Aarhus: \url{http://phys.au.dk/forskning/cscaa/}

\end{acknowledgements}

% WARNING
%-------------------------------------------------------------------
% Please note that we have included the references to the file aa.dem in
% order to compile it, but we ask you to:
%
% - use BibTeX with the regular commands:
%   \bibliographystyle{aa} % style aa.bst
%   \bibliography{Yourfile} % your references Yourfile.bib
%
% - join the .bib files when you upload your source files
%-------------------------------------------------------------------

\bibliographystyle{aa_url} % style aa.bst
\bibliography{main} % your references Yourfile.bib

\clearpage
\onecolumn

%--------------------------------------------------------------------
\begin{appendix} %First appendix

\section{Tests}
\subsection*{Multi-element non-LTE radiative transfer}\label{sect:A_multi_element}

\begin{figure}[h]%[h]
  \centering
  \resizebox{7.0in}{!}{\includegraphics{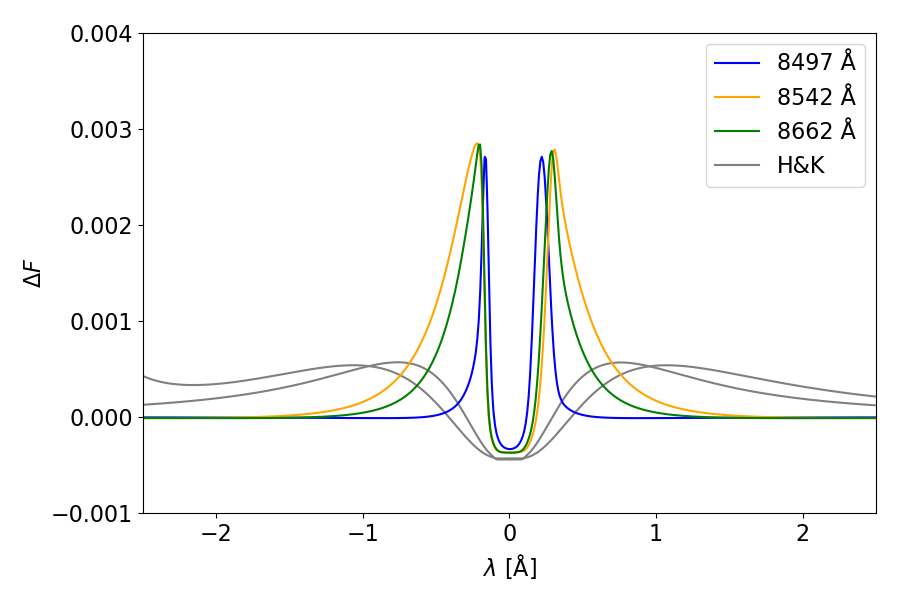}\includegraphics{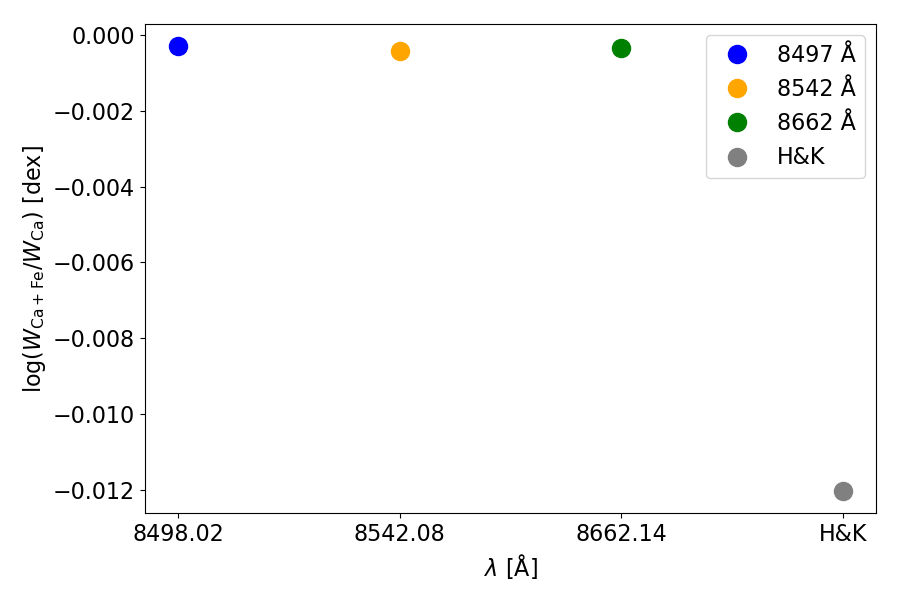}}
\caption[]{Difference in normalised flux and equivalent width between synthetic spectra with both Fe and Ca in non-LTE, and with only Ca in non-LTE. Computed in 1D for a metal-poor giant: $\teff=5000$ K, $\logg=2.0$ and $\feh=-2$.} 
\label{fig:CaFe_Ca}
\end{figure}

We investigated the effect of departures from 1D LTE for \ion{Fe}{I} on the \ion{Ca}{II} level populations for a metal-poor giant with stellar parameters: $\teff=5000$ K, $\logg=2.0$ and $\feh=-2$. The model atom describing \ion{Fe}{} that was used for these tests is described in \cite{Amarsi22}. In the reference `LTE' calculation, all Fe level populations were fixed to their LTE values while we allow \ion{Ca}{II} to depart from LTE. In the full 1D non-LTE calculation, both Fe and Ca are allowed to depart from LTE. We compare the full non-LTE spectra to the reference `LTE' spectra in \fig{fig:CaFe_Ca}. We find minimal effects on the line shape and equivalent width such that we can confidently keep Fe in LTE during our 3D non-LTE calculations.

\subsection*{Angle-dependence of spectrum synthesis}\label{sect:A_angledep}

\begin{figure}%[h]

  \centering
  \resizebox{7.0in}{!}{\includegraphics{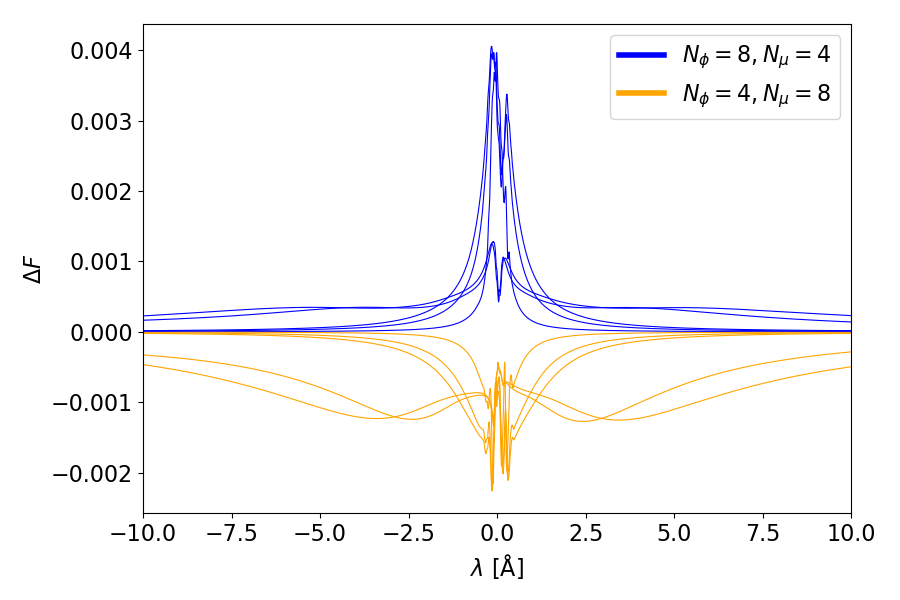}\includegraphics{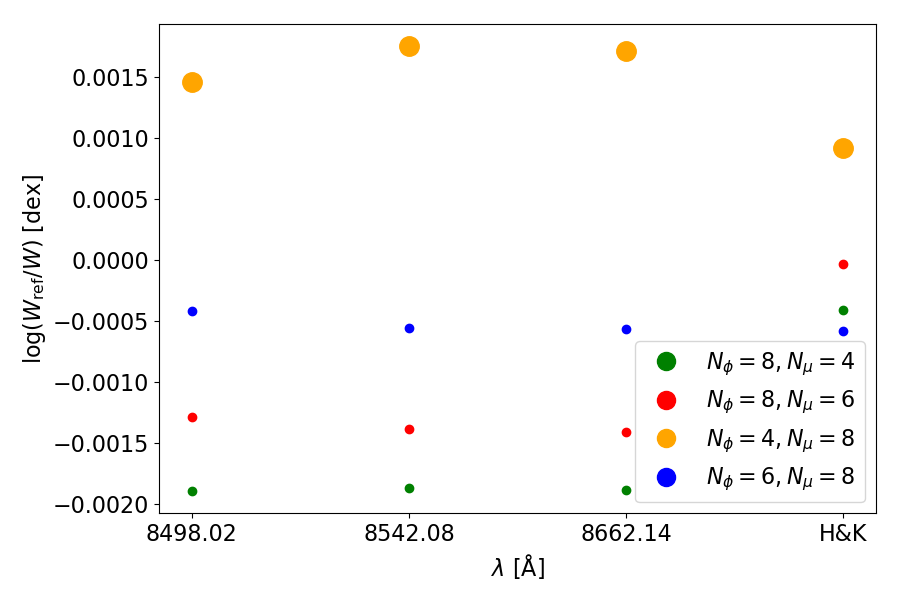}}
\caption[]{Difference in normalised flux and equivalent width between a reference spectrum with $N_\phi=8$ and $N_\mu=8$ rays and spectra with $N_\phi=4,6$ and $N_\mu=4,6$. The configuration used for the grid is denoted in orange. All spectra are computed in 3D non-LTE for a metal-poor giant: $\teff=5000$ K, $\logg=2.0$ and $\feh=-2$.} 
\label{fig:angles}
\end{figure}

\fig{fig:angles} shows the effect of number of $\phi$ and $\mu$ rays in the non-LTE iterations on the equivalent width for a metal-poor giant with stellar parameters: $\teff=5000$ K, $\logg=2.0$ and $\feh=-2$. The 3D non-LTE reference spectra has a configuration of $N_\phi=8$ and $N_\mu=8$. The final configuration used during the non-LTE iterations has $N_\phi=4$ and $N_\mu=8$, consistent with earlier work by the authors \citep{Amarsi18a,Amarsi22,Lagae23}. After the level populations have converged, the final emergent spectrum is computed using a configuration of $N_\phi=4$ and $N_\mu=6$.

\subsection*{Model resolution}\label{sect:A_modelres}

\begin{figure}%[h]
  \centering
  \resizebox{7.0in}{!}{\includegraphics{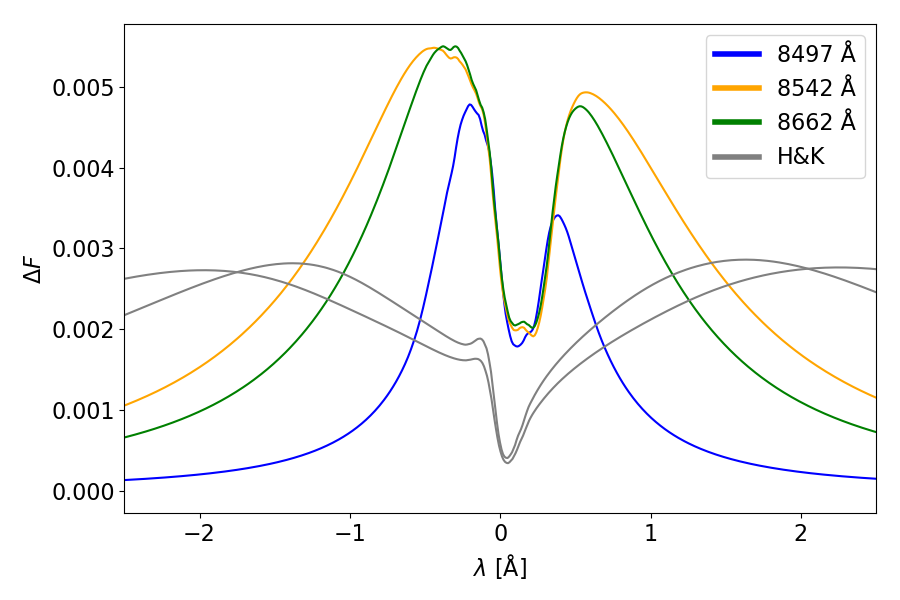}\includegraphics{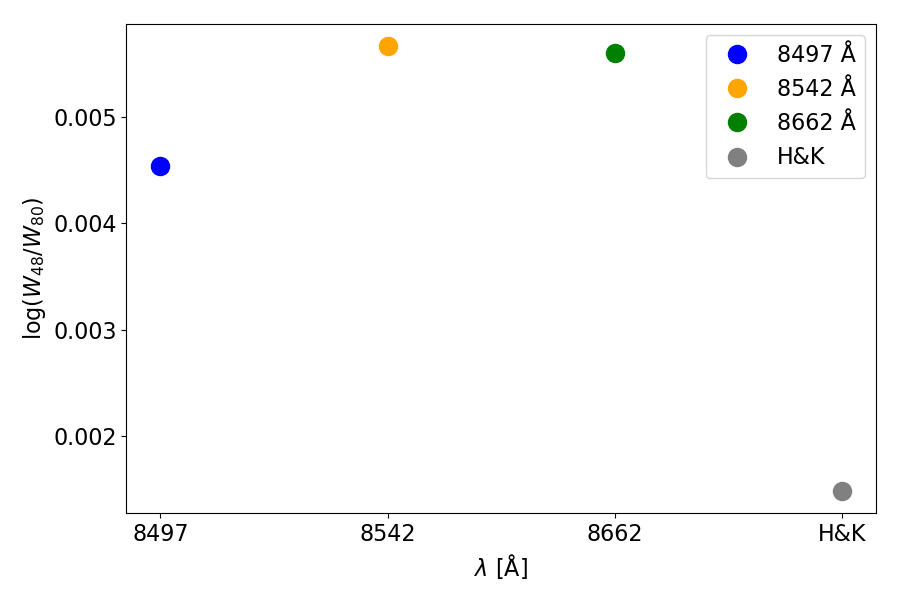}}
\caption[]{Difference in normalised flux and equivalent width between spectra computed with the same \stagger{} model, but different number of horizontal grid points: $N_\mathrm{hor}=80$ and $N_\mathrm{hor}=48$. All spectra are computed in 3D non-LTE for a metal-poor giant: $\teff=5000$ K, $\logg=2.0$ and $\feh=-2$.} 
\label{fig:resolution}
\end{figure}

\fig{fig:resolution} shows the effect of horizontal resolution on the equivalent width and line shape, in 3D non-LTE, for a metal-poor giant with stellar parameters: $\teff=5000$ K, $\logg=2.0$ and $\feh=-2$. Following \cite{Rodriguez24}, we investigated the impact of downsampling the horizontal mesh beyond 80 grid points. The final configuration used in this work has 48 grid points in the horizontal directions.

\subsection*{Temporal sampling}\label{sect:A_tempsamp}
\begin{figure}%[h]
  \centering
  \resizebox{7.0in}{!}{\includegraphics{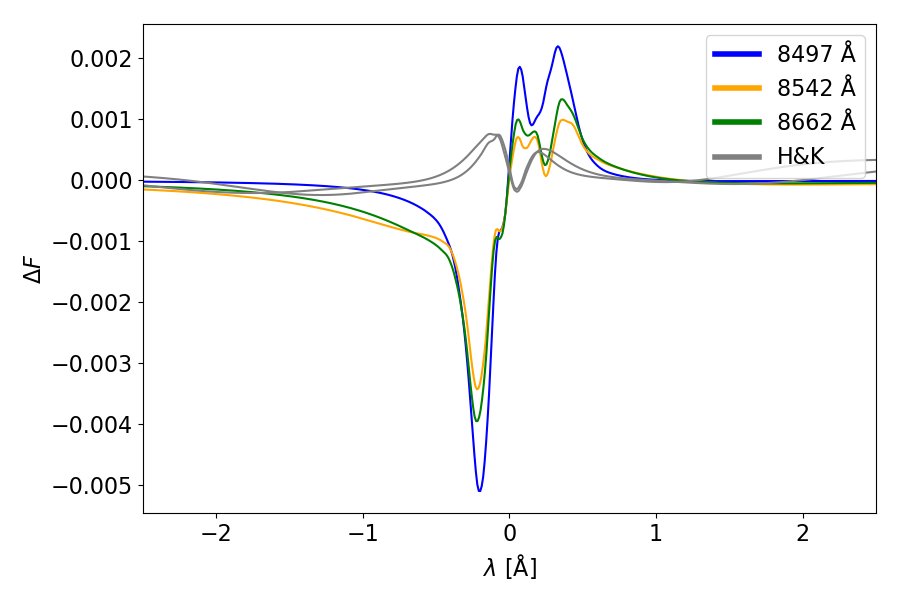}\includegraphics{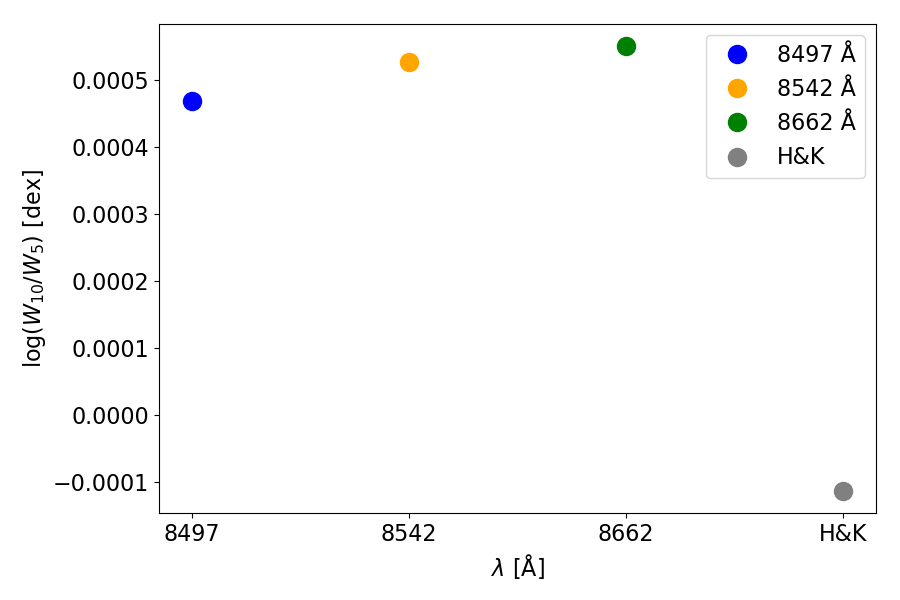}}
\caption[]{Difference in normalised flux and equivalent width between spectra computed with the same \stagger{} model, but different number of snapshots: $N_\mathrm{snap}=10$ and $N_\mathrm{snap}=5$. All spectra are computed in 3D non-LTE for a metal-poor giant: $\teff=5000$ K, $\logg=2.0$ and $\feh=-2$.} 
\label{fig:tempsamp}
\end{figure}

\fig{fig:tempsamp} shows the effect of temporal sampling on the equivalent width and line shape, in 3D non-LTE, for a metal-poor giant with stellar parameters: $\teff=5000$ K, $\logg=2.0$ and $\feh=-2$. Following \cite{Rodriguez24}, we investigated the impact of the number of snapshots. The final configuration used in this work uses five snapshots per model atmosphere.

%%%

\end{appendix}

\end{document}